\documentclass[transmag]{IEEEtran}

    \usepackage{amsmath,amssymb,amsfonts}
    \usepackage{mathtools}
    
    \usepackage{algorithmic}
    \usepackage{graphicx}
    \usepackage{textcomp}
    \usepackage{color}
    \usepackage[acronym,shortcuts]{glossaries}
    \usepackage[inline]{enumitem}

    \usepackage{placeins}
    \usepackage[keeplastbox]{flushend}
    \usepackage[utf8]{inputenc}
    \usepackage{tkz-kiviat,pgfplots}
    \pgfplotsset{compat=newest}
    \DeclareUnicodeCharacter{2212}{−}
    \usepgfplotslibrary{groupplots,dateplot}
    \usetikzlibrary{patterns,shapes.arrows}
    \usetikzlibrary{arrows}
    \usetikzlibrary{pgfplots.statistics}
    \usepackage{xcolor}
    \usepackage{mdframed}
    \usepackage{lipsum}
    \usepackage{booktabs}
    \usepackage[numbers]{natbib}
    \usepackage[binary-units=true]{siunitx}
    \DeclareSIUnit{\belmilliwatt}{Bm}
    \DeclareSIUnit{\dBm}{\deci\belmilliwatt}
    \usepackage{float}
    \usepackage[printwatermark]{xwatermark}

    \usepackage{pgfplots}
    \usepackage{setspace}
    \usepackage{subcaption}
    \usepgfplotslibrary{dateplot} 
    \usepackage{eurosym}
    \usepackage{tabularx}
    
    \usepackage{soul}
    
    \usepackage{filecontents}

    \usepackage{scalerel}
    \usetikzlibrary{svg.path}
    \usepackage{hyperref}
    \definecolor{orcidlogocol}{HTML}{A6CE39}
    \tikzset{
      orcidlogo/.pic={
        \fill[orcidlogocol] svg{M256,128c0,70.7-57.3,128-128,128C57.3,256,0,198.7,0,128C0,57.3,57.3,0,128,0C198.7,0,256,57.3,256,128z};
        \fill[white] svg{M86.3,186.2H70.9V79.1h15.4v48.4V186.2z}
                     svg{M108.9,79.1h41.6c39.6,0,57,28.3,57,53.6c0,27.5-21.5,53.6-56.8,53.6h-41.8V79.1z M124.3,172.4h24.5c34.9,0,42.9-26.5,42.9-39.7c0-21.5-13.7-39.7-43.7-39.7h-23.7V172.4z}
                     svg{M88.7,56.8c0,5.5-4.5,10.1-10.1,10.1c-5.6,0-10.1-4.6-10.1-10.1c0-5.6,4.5-10.1,10.1-10.1C84.2,46.7,88.7,51.3,88.7,56.8z};
      }
    }
    \newcommand\orcidicon[1]{\href{https://orcid.org/#1}{\mbox{\scalerel*{
    \begin{tikzpicture}[yscale=-1,transform shape]
    \pic{orcidlogo};
    \end{tikzpicture}
    }{|}}}}

    \definecolor{colorCE0}{HTML}{fc8d59}
    \definecolor{colorDarkCE0}{HTML}{D67147}
    \definecolor{colorCE1}{HTML}{99d594}
    \definecolor{colorDarkCE1}{HTML}{628F5E}
    \definecolor{colorCE2}{HTML}{3288bd}
    \definecolor{colorDarkCE2}{HTML}{286FA2}

    \definecolor{colorSF7}{HTML}{1b9e77}
    \definecolor{colorSF8}{HTML}{d95f02}
    \definecolor{colorSF9}{HTML}{7570b3}
    \definecolor{colorSF10}{HTML}{e7298a}
    \definecolor{colorSF11}{HTML}{66a61e}
    \definecolor{colorSF12}{HTML}{e6ab02}

    \definecolor{colorLoRa}{HTML}{af8dc3}
    \definecolor{colorNBIoT}{HTML}{7fbf7b}
    \definecolor{colorMultiRAT}{HTML}{fc8d59}

    \newcommand{\tkzKiviatLegend}[3]{\begin{tikzpicture}
        \draw [thick, colorLoRa] (0,0) -- (0.5,0); 
        \node at (1.75,0) {#1};
        \draw [thick, colorNBIoT] (2.75,0) -- (3.25,0); 
        \node at (4,0) {#2};
        \draw [dashed, thick, colorMultiRAT] (0,-0.50) -- (0.5,-0.50); 
        \node[anchor=west] at (0.8,-0.5) {#3};
    \end{tikzpicture}}

    \makeglossaries
    \newacronym{gprs}{GPRS}{General Packet Radio Services}
    \newacronym{egprs}{EGPRS}{Enhanced Data Rates for GSM Evolution}
    \newacronym{iot}{IoT}{Internet of Things}
    \newacronym{nbiot}{NB-IoT}{Narrowband IoT}
    \newacronym{lte}{LTE}{Long Term Evolution}
    \newacronym[plural=LPWANs,firstplural=Low Power Wide Area Networks (LPWANs)]{lpwan}{LPWAN}{Low Power Wide Area Network}
    \newacronym{lpwa}{LPWA}{Low Power Wide Area}
    \newacronym{lorawan}{LoRaWAN}{Long Range Wide Area Network}
    \newacronym{lora}{LoRa}{Long Range}
    \newacronym{nb}{NB}{Narrowband}
    \newacronym{m2m}{M2M}{Machine to Machine}
    \newacronym{cbm}{CBM}{Condition Based Maintenance}
    \newacronym{ai}{AI}{Artificial Intelligence}
    \newacronym{edrx}{eDRX}{Extended Discontinuous Reception Mode}
    \newacronym{drx}{DRX}{Discontinuous Reception Mode}
    \newacronym{psm}{PSM}{Power Saving Mode}
    \newacronym{gps}{GPS}{Global Positioning System}
    \newacronym{ptw}{PTW}{Paging Time Window}
    \newacronym{tau}{TAU}{Tracking Area Update}
    \newacronym{mtc}{MTC}{Machine-Type Communication}
    \newacronym{ue}{UE}{User Equipment}
    \newacronym{cdrx}{CDRX}{Connected Mode DRX}
    \newacronym{3gpp}{3GPP}{3rd Generation Partnership Project}
    \newacronym{gsm}{GSM}{Global System for Mobile Communications}
    \newacronym{rf}{RF}{Radio Frequency}
    \newacronym{prbs}{PRBs}{Physical Resource Blocks}
    \newacronym{mcl}{MCL}{Maximum Coupling Loss}
    \newacronym{ce}{CE}{Coverage Enhancement}
    \newacronym{rsrp}{RSRP}{Reference Signals Received Power}
    \newacronym{toa}{ToA}{Time on Air}
    \newacronym{rrc}{RRC}{Radio Resource Connection}
    \newacronym{rtc}{RTC}{Real Time Clock}
    \newacronym{css}{CSS}{Chirp Spread Spectrum}
    \newacronym{ism}{ISM}{Industrial, Scientific and Medical}
    \newacronym{mmtc}{MMTC}{Massive Machine Type Communication}
    \newacronym{sf}{SF}{Spreading Factor}
    \newacronym{cr}{CR}{Coding Rate}
    \newacronym{bw}{BW}{Bandwidth}
    \newacronym{adr}{ADR}{Adaptive Data Rate}
    \newacronym{abp}{ABP}{Authentication By Personalisation}
    \newacronym{otaa}{OTAA}{Over The Air Authentication}
    \newacronym{ofdm}{OFDM}{Orthogonal Frequency Division Multiplexing}
    \newacronym{scfdma}{SCFDMA}{Single-Carrier Frequency Division Multiple Access}
    \newacronym{udp}{UDP}{User Datagram Protocol}
    \newacronym{qos}{QoS}{Quality of Service}
    \newacronym{ra}{RA}{Random Access}
    \newacronym{rar}{RAR}{Random Access Response}
    \newacronym{snr}{SNR}{Signal-to-Noise Ratio}
    \newacronym{ota}{OTA}{Over The Air}
    \newacronym{ttm}{TTM}{Time To Market}
    \newacronym{cots}{COTS}{commercial off-the-shelf}
    \newacronym{mac}{MAC}{Medium Access Control}
    \newacronym{multi-rat}{Multi-RAT}{Multiple Radio Access Technology}
    \newacronym{rat}{RAT}{Radio Access Technology}
    \newacronym{pcb}{PCB}{Printed Circuit Board}
    
    \setcitestyle{square}
    \def\BibTeX{{\rm B\kern-.05em{\sc i\kern-.025em b}\kern-.08em
        T\kern-.1667em\lower.7ex\hbox{E}\kern-.125emX}}

    \newcommand{\lievenok}[1]{}
    \newcommand{\liesbetok}[1]{}
    \newcommand{\gillesok}[1]{}
    \newcommand{\geofok}[1]{}
    \newcommand{\guusok}[1]{}
    
    \newcommand{\new}[1]{{#1}}
    \newcommand{\old}[1]{{}}
    \newcommand{\technology}[1]{{\textbf{\textit{#1}-- }}}
    \newcommand{\nbiot}{{\technology{\acrshort{nbiot}}}}
    \newcommand{\lorawan}{{\technology{\acrshort{lorawan}}}}
    
    \makeatletter
    \def\ps@IEEEtitlepagestyle{%
    \def\@oddfoot{\mycopyrightnotice}%
    \def\@evenfoot{}%
    }
    \def\mycopyrightnotice{%
    {\footnotesize This work has been submitted
    to the IEEE for possible publication. \\Copyright may be transferred
    without notice, after which this version may no longer be accessible.} 
    \gdef\mycopyrightnotice{}
    }

\begin{document}

\title{Multi-RAT for IoT: The Potential in Combining LoRaWAN and NB-IoT}

\author{
    \IEEEauthorblockN{Guus Leenders \orcidicon{0000-0001-9633-2584}, Gilles Callebaut \orcidicon{0000-0003-2413-986X}, Geoffrey Ottoy \orcidicon{0000-0002-3690-2458}, Liesbet Van der Perre \orcidicon{0000-0002-9158-9628}, Lieven De Strycker \orcidicon{0000-0001-8172-9650}\\}
    \IEEEauthorblockA{
        \textit{KU Leuven, ESAT-DRAMCO, Ghent Technology Campus}\\
        Ghent, Belgium\\
        name.surname@kuleuven.be
    }
}
\IEEEtitleabstractindextext{\begin{abstract}

The broad range of requirements of \acrlong{iot} applications has lead to the development of several dedicated communication technologies, each tailored to meet a specific feature set. A solution combining different wireless technologies in one device, can overcome the disadvantages of any individual technology. The design of such \acrlong{multi-rat} solutions based on the diverse characteristics of the technologies offers interesting opportunities. \new{We have assessed both the potential gains and the overhead that a \gls{multi-rat} solution brings about.}
To that end, we have evaluated key \acrshort{iot} node requirements in function of payload size and link quality: 
 energy efficiency,  coverage, payload size, 
 latency performance,
 \acrlong{qos}, and
 cost efficiency.
Our \new{assessment} and experimental validation of these features show the merits of a \gls{multi-rat} solution. Notably, energy consumption in use cases with only sporadic large payload requirements, can be improved by a factor of at least~4 with respect to either single-mode technologies. Moreover, latency-critical messages can get delivered on time and coverage can be extended elegantly where needed. 

\end{abstract}

}

\maketitle

\section{Introduction}

The \acrlong{iot} (\acrshort{iot}) is a key technological component in ever more applications. \glspl{lpwan}, a subset of IoT connectivity solutions, enable connecting objects over long distances. For example, trees equipped with low-power wireless sensors, allow for remote monitoring of the trees' health
, and in a more urban setting, low-power sensors create smart buildings and cities.

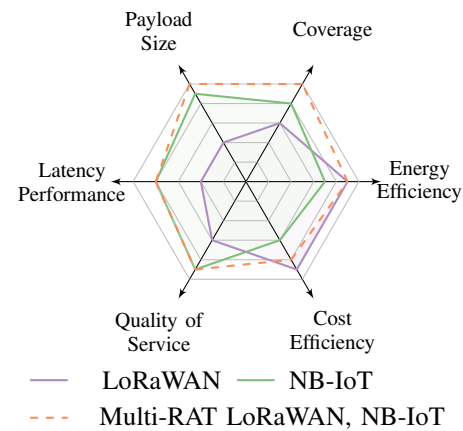
\begin{figure}[tbp]
    \centering
    \begin{tikzpicture}
\tkzKiviatDiagram[scale=0.3,label distance=.2cm,
        radial  = 6,
        gap     = 1,  
        label space=2.2,
        label style/.append style={font=\footnotesize},
        lattice = 5.5]{{Energy\\Efficiency}, {Coverage}, {Payload\\Size},{Latency\\Performance}, {Quality of\\Service}, {Cost \\Efficiency}}
\tkzKiviatLine[thick,color=colorLoRa,mark=none,   fill=colorLoRa!20,opacity=.5](4.5,3,2,2,3,4.5) 
\tkzKiviatLine[thick,color=colorNBIoT,                      fill=colorNBIoT!20,opacity=.5](3.5,4,4.5,4,4.5,3) 
\tkzKiviatLine[dashed, thick,color=colorMultiRAT,                      fill=colorMultiRAT!0,opacity=.5](4.5,5,5,4,4.5,4) 
 
\end{tikzpicture}

\tkzKiviatLegend{LoRaWAN}{NB-IoT}{Multi-RAT LoRaWAN, NB-IoT}
    \caption{\small Comparative study of the explored \gls{iot} network technologies, presenting the main \gls{iot} feature requirements.  }
    \label{fig:loravsnbiot}
\end{figure}

When matching wireless connectivity with the requirements of a specific \gls{iot} application, both technical and non-technical aspects should be considered, as illustrated in Fig.~\ref{fig:loravsnbiot}. Although each application has its own specific requirements, a large class requires a similar set of features such as long-range wireless connectivity, low energy consumption and cost effectiveness~\cite{mekki2018overview} while only transmitting a relatively small amount of data.

Conventional cellular standards (3G, 4G) have been designed to provide global coverage, yet they consume too much energy for battery-powered devices~\cite{balani2007energy}. The introduction of \glspl{lpwan} has enabled developers to opt for a cost effective low-power connectivity technology, whilst still enabling long-range communication.
\Gls{lorawan} and \gls{nbiot} are two prominent \gls{lpwan} technologies, operating in unlicensed and licensed bands respectively.

\Gls{lorawan} is a network stack that is implemented on top of the \gls{lora} physical layer.
It has been rolled out by both commercial operators and non-profit organisations, resulting in wide scale \gls{lorawan} coverage. 

\gls{nbiot} is  a cellular \gls{lpwan} technology. By reducing the \gls{ue}'s complexity and deploying a dedicated modulation inherited from \gls{lte}, its battery life and coverage are greatly extended, at the cost of reduced bandwidth and downlink modes. In contrast to traditional cellular transmission, 
\gls{nbiot} is optimized for \gls{lpwan} communication: providing a relatively small payload size and long-range communication. 

\new{For many \gls{iot} applications, a single technology suffices. However, use cases under varying operating conditions, may benefit from the integration of multiple wireless technologies.} Integration of multiple wireless technologies on a single device can lead to improved performances. Such \Acf{multi-rat} strategies have previously been used for short-range \gls{iot} devices, e.g., incorporating both Wi-Fi and Bluetooth connectivity~\cite{lee2015transfer}. Similarly, in cellular networks, different technologies are being used in a fall-back strategy. 
Unlicensed \gls{iot} technologies such as Sigfox and LoRaWAN have been combined into a single modem module. However, a true  \gls{multi-rat} approach with dynamic switching between both unlicensed and cellular technologies is lacking.

Specifically for battery-operated \gls{iot} applications, a \gls{multi-rat} solution can lead to longer autonomy, since the wireless transmission represents the main energy cost~\cite{karl2007protocols}. Based on the network conditions or the current needs, e.g., low latency or larger payload size, the \gls{iot} device can dynamically switch to the most \new{suitable technology, whilst optimizing for energy efficiency. }

\new{This particularly benefits more advanced, dynamic use cases where multiple types of packets need to be sent, or where coverage is not guaranteed. }
For example, a microphone enabled \gls{iot} sensor could send periodic sound level readings using a  technology that is most energy efficient for small payloads. When certain events are detected, the device could switch to a connectivity technology that is better suited to transmit a longer sound sample for classification in the cloud.

\technology{Contributions} The main contribution of this paper is a multi-RAT solution that improves on the performance parameters of LPWAN connections, and in particular can prolong the autonomy of battery-powered \gls{iot} devices. We focus on the wireless link between the device and the gateway, acknowledging that the operation of the gateways back-end infrastructure will also have an impact on feature requirements such as \gls{qos} and cost efficiency. In the design of a \gls{multi-rat} solution, the typical requirements and feature set of the \gls{iot} technologies are examined. We map these for both \gls{lorawan} and \gls{nbiot} and provide concrete figures that will be at the basis of the dynamic switching between different \gls{lpwan} standards.
Although both \gls{lorawan} and \gls{nbiot} are designed for low-power communication, measurements have shown that large differences in a node's power consumption can be experienced~\cite{leenders2020experimental}.

In~\cite{mikhaylov2018multi}, the authors have studied the energy consumption of both \gls{lorawan} and \gls{nbiot} for a smart city application and identify possible opportunities when using \gls{multi-rat}.

The specific contributions of this work are twofold. First, 
we determine the \new{specification-based}  boundaries of the most important properties such as coverage, payload size and energy efficiency.  The second contribution is the experimental validation of the Multi-RAT potential integrating two \gls{lpwan} technologies: \gls{lorawan} and \gls{nbiot}. 
We specifically show that for both energy efficiency and latency, a \gls{multi-rat} solution outperforms single technology solutions in scenarios with dynamic payload size, link quality, or \gls{qos} requirements. 

\new{Yet, we also clarify some arguments against deploying a \gls{multi-rat} system.}  

This paper is organized as follows. We determine the theoretical boundaries of a \gls{multi-rat} system, set by the individual \gls{lpwan} specifications, exploring both \gls{nbiot} and \gls{lorawan}. Secondly, our measurements are presented in order to evaluate the energy efficiency and latency in real-life situations. Finally, we summarize our main findings and show how a fitting wireless technology can be selected in a dynamic, \gls{multi-rat} application.

\section{LPWAN Multi-RAT support for IoT Requirements}

We analyze the main theoretical  properties of \gls{lorawan} and \gls{nbiot}, illustrated in Fig.~\ref{fig:loravsnbiot}, 
thereby assessing the potential \gls{multi-rat} opportunities.

\subsection{Energy Efficiency}

Both  \gls{lorawan} and \gls{nbiot} provide measures to minimize energy in support of battery-powered devices. A \gls{multi-rat} system for \gls{iot} should select the most energy efficient \gls{iot} technology within a set of application parameters (e.g., payload, latency toleration). 

\lorawan
\Gls{lorawan} is low-power by design. The simple \gls{mac} and narrow bandwidth signals contribute to a reduced power consumption. When utilizing \gls{lorawan}, the energy consumption is mainly determined by the \gls{toa} which depends on (I) the payload size and (II) the \acrlong{sf}.

\new{One of the advantages of implementing \acrlong{adr} in the \gls{lorawan} specification, is optimizing energy efficiency.} 
This algorithm alters the transmit power and data rate of the end-devices depending on the wireless channel conditions. In adverse scenarios, the transmit power and/or spreading factor is increased in order to overcome a low \gls{snr}. \new{In contrast, when the received \gls{snr} is higher than necessary, the device will decrease its transmit power and/or spreading factor.} 

\nbiot
To address typical \gls{lpwan}-specific requirements, the \gls{lte} standard was simplified with an emphasis on energy reduction and lower complexity. Most sleep timers have been extended in the release of \gls{nbiot} in comparison with \gls{lte}. In this manner the end-devices can sleep for a longer time period before making contact to the network.
The necessity to monitor paging, i.e., listening for down link messages, is drastically reduced by introducing \gls{edrx}. This reduces active radio time, lowering the energy footprint. 
Long \gls{edrx} cycles allow the \gls{nbiot} node to sleep for \new{up to \SI{186}{minutes}}, after which the node checks for paging before going to sleep again. 

When a node no longer needs to maintain an active connection, it can go into \gls{psm}. In this mode, the node completely disconnects from the network.  
For a detailed overview of energy consumption in different stages, we refer to Section~\ref{sec:evaluation_energy}.
\lievenok{moeten we niet vanaf hier al de vergelijking maken vanuit het Multi-RAT perspectief ? of toch zeker vanaf de experimentele metingen ?}
\lievenok{Misschien te vergaand: II en III samenvoegen, maw de theoretische bespreking direct laten volgen door de metingen en dan direct het Multi-RAT perspectief erbij zetten ?}

\subsection{Coverage}
When deploying \gls{iot} nodes in a city landscape or in remote locations, good coverage is critical for obtaining a reliable communication. Both \gls{lorawan} and \gls{nbiot} feature mechanisms to improve range and coverage. The distance to the nearest gateway clearly is the main determining factor for reliability for both technologies. The networks typically do not share a common infrastructure. Hence, there is inherent redundancy in a \gls{multi-rat} solution and the reliability can be greatly improved with respect to any single \acrshort{rat} \gls{iot} connectivity.

\lorawan
The range strongly depends on the utilized data rate. Increasing the spreading factor lowers the required demodulation floor and thereby extends the range. 
The \gls{lora} link budget can be as high as \SI{156}{\decibel}~\cite{wixted2016evaluation} (\acrshort{sf} 12).

In contrast to cellular networks, private networks can be deployed in \gls{lorawan}. Therefore, the coverage can be extended by adding more gateways to the network.

\nbiot
Three \gls{ce} levels are baked into \gls{nbiot}. They enable providers to provide network connectivity in places that are hard to reach, such as subterranean parking lots. This way, the \gls{mcl} is improved by \SI{20}{\decibel}, compared to  \gls{gprs}:  up to \SI{164}{\deci\bel}. Uplink messages are typically transmitted at \SI{23}{\deci\belmilliwatt} \cite{andres2017narrowband}.

\gls{ce} levels are decided upon based on the quality of the \gls{rf} channel, estimated by the \gls{rsrp}: the power of the Reference Signals spread over the full bandwidth. During the network random access procedure, two \gls{rsrp} thresholds are configured. Therefore, three \gls{ce} levels can be configured: \gls{ce} level~0 (good signal quality), 1 (mediocre signal quality) and 2 (poor signal quality). If there is poor network reception when establishing a connection, the \gls{ce} level will be increased to level~1~or~2. High \gls{ce} levels result in lower bandwidth and packet repetitions~\cite{liberg2017cellular}: increasing energy consumption by increased \gls{toa}.

\new{\mbox{T-Mobile}~\cite{tmobile} performed a measurement campaign to study the distribution of required \gls{ce} levels in real-life conditions. }
No coverage extension was enforced outdoors for \SI{93}{\percent} of measurements. For subterranean locations, \gls{ce} level 1 is used \SI{27}{\percent}, while \gls{ce} level~2 is used \SI{19}{\percent} of the time. 

\begin{figure*}[t!]
	\centering
	\begin{subfigure}[t]{0.49\linewidth}
    \centering
    \input{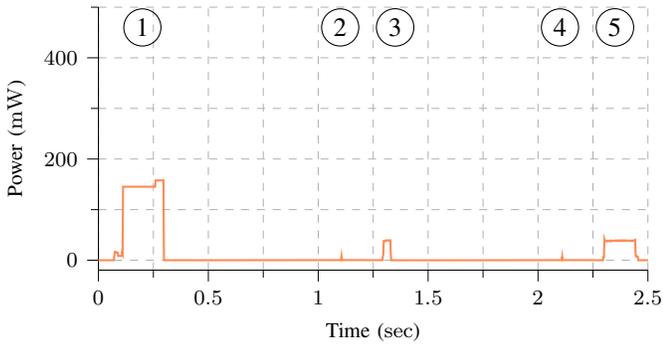}
    \caption{\small Measured power consumption of a \gls{lora} node (at \gls{sf} 9): (1) transmit, (2) processing, (3) first receive window, (4) processing, (5) second receive window. Network registration is omitted, as this is not obligatory in \gls{lorawan}.}  
    \label{fig:currentdrawlora}
	\end{subfigure}\hfill
	\begin{subfigure}[t]{0.49\linewidth}
    \centering
    \input{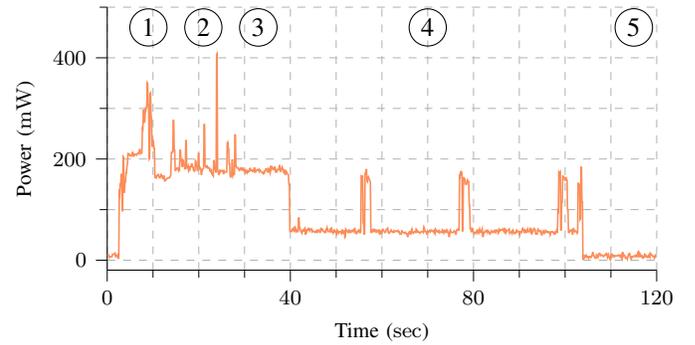}
    \caption{\small Measured power consumption of a \gls{nbiot} node (at \gls{ce} level~0): (1) network search and join, (2) package sending, (3) \acrlong{cdrx}, (4) \acrlong{edrx}, (5) \acrlong{psm}. Network registration is obligatory and is only performed once (as long as power is maintained). }
    \label{fig:currentdrawnbiot}
	\end{subfigure}
	\caption{\small Experimentally determined power consumption of \gls{lora} and \gls{nbiot} side by side. Note the large difference in timing, resulting in a larger energy consumption for \gls{nbiot}. }\label{fig:currentdraw} 
\end{figure*}

\subsection{Payload Size}
Typical \gls{iot} nodes require only a limited amount of bytes to be sent. However, a more generous payload size limit will allow a node to periodically send larger packets. This is particularly useful in surveillance use cases (uplink) and \gls{ota} updates (downlink).

Implementing a \gls{multi-rat} solution provides the combined range of available payload sizes. 

\lorawan
The payload size defined in \gls{lorawan} depends on the utilized data rate.
For higher spreading factors, \gls{lorawan} restricts the payload size to 51 bytes, while in more optimal conditions a \SI{242}{\byte} payload can be used in one 
message (both in uplink and downlink). 
In case a longer payload is required, the developer has to divide the payload over multiple messages, thereby increasing the reception latency and energy consumption.
        
\nbiot
The maximum payload size for each message, regardless of conditions, is \SI{1600}{\byte}~\cite{mekki2019comparative}, both for uplink and downlink messages. There are no  duty cycle limits in \gls{nbiot}: an unlimited amount of messages can be sent per day.

\subsection{Latency}
We consider two latency requirements, i.e. uplink and downlink latency. Uplink latency is defined as the time between the intent of sending a message on the device and receiving that message on the server. This latency, in practice,  can be largely attributed to the latency between the \gls{iot} node and the gateway. Notably, there is a considerable difference between the intent of sending a message and actual sending the message in license-exempt bands. Downlink latency is defined as latency in the reverse direction. 

\new{By characterizing the latency of the implemented technologies, an ideal technology can be selected for latency-critical packets. As such, the uplink latency is reduced to that of the fastest technology. This emphasises the need for a fine-grained profile of these technologies, as illustrated in this work.}

\lorawan
The latency of the downlink messages is constrained by the adopted device classes, while the latency of the uplink messages is limited by the duty cycle and data rate.

\begin{itemize}
    \item Due to license-exempt operation, the devices are subjected to a maximum duty cycle\liesbetok{kunnen we vermelden wat deze/range is om het minder abstract te maken?} (1\% for EU \SI{868}{\mega\hertz}). This could increase the latency of uplink \lievenok{deze afkorting wordt blijkbaar alleen hier gebruikt, verder wordt het steeds voluit geschreven}messages as they have to respect a minimum wait time before transmitting the next message. 
    \item 
    Three classes of \gls{lorawan} devices have been defined: A, B and C. The classes vary on different scheduling for down link messages. Class A devices feature two down link windows after each up link message. Class B adds time synchronized receive windows in between up link packets. Class C devices are constantly able to receive packets as their receive window remains open unless they are transmitting data. Consequently, application requiring low latency downlink messages should adopt class C.
 
Furthermore, the data rate will have an impact on the uplink latency.
The data rate of \gls{lorawan} depends on the utilized \gls{sf}, bandwidth and \gls{cr}. The time on air of a \gls{lorawan} package, ranges from  \SI{25}{\milli\second} (\gls{sf}7, \SI{1}{\byte}) to \SI{2.5}{\second} (\gls{sf}12, \SI{51}{\byte}).
\end{itemize}

\nbiot
While uplink latency is limited to \SI{10}{\second} at most~\cite{ratasuk2016overview} in \gls{nbiot}, downlink latency heavily depends on the set timer values \gls{edrx} and \gls{psm}. 
\begin{itemize}
    \item Uplink latency is mainly influenced by the path loss and the deployment method of the base station~\cite{liberg2017cellular}. 
    An extensive \gls{nbiot} latency model has been documented by \citeauthor{azari2018latency} \cite{azari2018latency}. When a device needs to send an uplink message, it first needs to listen for cell information. Through this information gathering, the node synchronises with the base station. By sending a \gls{ra} request to the base station, the device performs access reservation. The base station responds by sending a \gls{rar}, indicating resources reserved for the \gls{nbiot} transfer. Finally, the device is able to send data to the base station. 

    \item Both \gls{edrx} timers and \gls{psm} timers regulate when the node is able to receive data, thus controlling downlink latency. By prolonging the \gls{edrx} cycle, more periodic paging cycles will occur. When in \gls{psm}, no packets can be received until the \gls{tau} message is sent.
\end{itemize}

Typical latency figures reported in literature range from \SI{0.3}{\second} to \SI{8.3}{\second}, depending on link budget and deployment type~\cite{liberg2017cellular}.
In a stand-alone deployment scenario, full base station power is available to \gls{nbiot},  improving latency. In good coverage conditions, latency is predominantly caused by the time to acquire synchronization and waiting for an access opportunity. Latency in poor coverage conditions are generally caused by latency of the exception report. \gls{nbiot} features a down link data rate of maximal \SI{200}{kbps} and uplink of maximal \SI{180}{kbps}~\cite{liberg2017cellular}.

\subsection{Quality of Service}
In IoT-based monitoring use cases, \gls{qos} focuses on packet loss and throughput.  
Mechanisms such as enabling repetitions, increasing output power, etc. allow technologies to control \gls{qos} dynamically. 

\lorawan The increased interference due to license-exempt operation, is addressed by employing the spread spectrum technique \gls{css}. \Gls{lorawan} does not support different \gls{qos} levels. However, an acknowledgment can be requested so the device can retransmit a message if lost.  

\nbiot
As \gls{nbiot} operates in licensed spectrum, it can offer greater \gls{qos} than networks operating in an unlicensed spectrum. Furthermore, \gls{nbiot} employs the same proven time slotted synchronous protocol as used in \gls{lte}: ensuring end-to-end \gls{qos}.

\subsection{Cost Efficiency}
Evidently, the initial hardware investment cost of a Multi-RAT solution will be higher than for a single radio. Several costs need to be considered to get the total cost of ownership: 
spectrum cost, network deployment cost, the cost of the end device, and eventual costs related to replacement or recharging of batteries involving service visits (relating to energy efficiency). 
 
\lorawan As \gls{lorawan} operates in the unlicensed bands, no spectrum cost is applicable. However, to extend range and network manageability, private networks can be deployed.

\nbiot
Both the spectrum license and base station contribute significantly to the combined cost of a \gls{nbiot} network, due to \gls{nbiot} running in licensed spectrum. To illustrate, the cost of acquiring the necessary spectrum can amount to \EUR{500M} per \si{\mega\hertz}~\cite{mekki2019comparative}.

\section{Experimental Evaluation\label{sec:experimental_evaluation}}
\gls{lorawan} and \gls{nbiot} have been thoroughly compared and documented in literature; however, experiments-based comparisons of both \emph{energy consumption} and \emph{latency} are lacking. In what follows, these parameters are validated in the field. 

\subsection{Experiment Setup}
To evaluate the aforementioned \gls{iot} feature requirements, custom hardware has been developed for both \gls{lora} and \gls{nbiot} measurements, using \gls{cots} electronic components. All files are open source~\cite{source} to facilitate reproducibility. \new{In order to assess the technologies in real-life conditions, including the impact of the network-specific configurations, the Proximus network infrastructure is used \textit{as-is}.}

\lorawan Our \gls{lorawan} experiments were conducted with a Happy Gecko starter kit, running the \gls{lorawan} network stack, connected to a \gls{lora} extension module. 
The board hosts a Semtech SX1272 \gls{lora} Radio chip and runs the \gls{lorawan} stack. The energy profile of the system is measured with the built-in power monitor of the Happy Gecko.

\nbiot
The \gls{nbiot} measurements presented in this paper were performed using a custom-made implementation of the Quectel BG96 module. The BG96 is used in a setup which allows us to measure the energy consumption per modem state of the \gls{nbiot} node. 

\begin{figure}[tbp]
    \centering
    \input{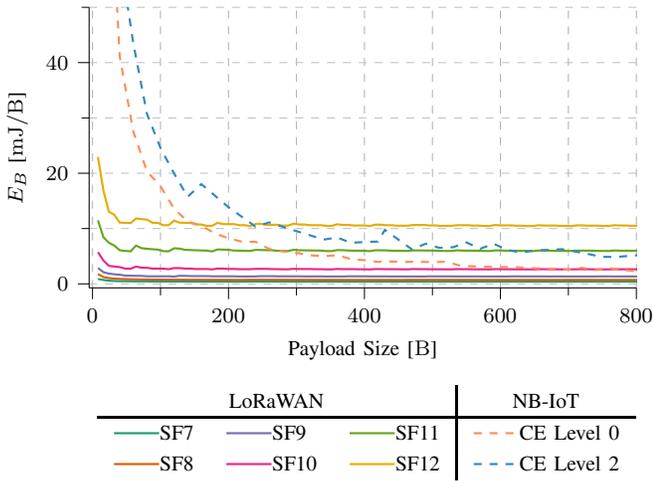}
    \caption{\small Energy consumption per byte comparison between \gls{lorawan} and \gls{nbiot}. The difference in energy consumption is most noticeable when sending small payloads. Over \new{\SI{240}{\byte}}, \gls{nbiot} \new{(\gls{ce} level 2)} is more energy efficient per byte, with respect to \gls{lorawan} (\gls{sf} 12). 
    }
    \label{fig:energyperbyte}
\end{figure}

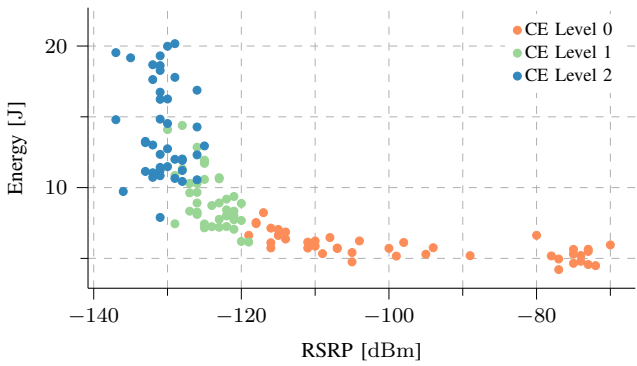
\begin{figure}[tbp]
    \centering
\begin{tikzpicture}

\colorlet{color1}{colorCE0}
\colorlet{color4}{colorCE1}
\colorlet{color5}{colorCE2}

\begin{axis}[%
axis lines* = {left},
tick align=outside,
tick pos=left,
width = \linewidth,
height = 0.6\linewidth,
separate axis lines,
x grid style={white!69.01960784313725!black},
grid,
grid style={help lines,color=gray!50, dashed},
xmin=-140.77178, xmax=-66.62358,
xtick style={color=black},
y grid style={white!69.01960784313725!black},
ytick={0,10,20},
xtick={-140,-120,-100,-80,-60},
ymin=2.89056, ymax=22.70769,
ytick style={color=black},
legend style={draw=none, nodes={scale=0.7, transform shape}},
xlabel={RSRP [\si{\dBm}]},
ylabel={Energy [\si{\joule}]},
label style={font=\footnotesize},
tick label style={font=\footnotesize},
minor tick num=1,
yminorgrids,
ymajorgrids,
xminorgrids
]
\addplot [only marks, mark=*, mark size=1.5pt, draw=color1, fill=color1, colormap/viridis]
table{%
x                      y
-119.00000 6.62973
-118.00000 7.45417
-118.00000 7.53283
-117.00000 8.23190
-116.00000 6.10285
-116.00000 7.13637
-116.00000 5.73844
-115.00000 7.04443
-115.00000 6.62073
-114.00000 6.36626
-114.00000 6.86505
-111.00000 5.72806
-111.00000 6.14779
-110.00000 5.86173
-110.00000 6.22797
-109.00000 5.33608
-108.00000 6.47111
-107.00000 5.70141
-107.00000 5.73046
-105.00000 4.75373
-105.00000 5.41399
-104.00000 6.23227
-100.00000 5.71132
-99.00000 5.16229
-98.00000 6.12374
-95.00000 5.29644
-94.00000 5.75344
-89.00000 5.19242
-80.00000 6.62887
-78.00000 5.17348
-77.00000 4.20855
-77.00000 4.95602
-75.00000 4.65286
-75.00000 5.29805
-75.00000 5.63507
-74.00000 4.79664
-74.00000 5.20403
-73.00000 5.64619
-73.00000 4.56572
-73.00000 5.49128
-72.00000 4.48788
-70.00000 5.95012
};
\addlegendentry{CE Level 0}

\addplot [only marks, mark=*, mark size=1.5pt, draw=color4, fill=color4, colormap/viridis]
table{%
x                      y
-130.00000 14.10063
-129.00000 10.85486
-129.00000 7.44360
-128.00000 14.38983
-128.00000 11.32614
-127.00000 10.30209
-127.00000 9.63297
-127.00000 8.33477
-126.00000 9.66211
-126.00000 8.23243
-126.00000 8.10336
-126.00000 8.91433
-126.00000 8.26592
-126.00000 12.85321
-126.00000 10.29418
-125.00000 7.15848
-125.00000 10.58148
-125.00000 11.87369
-125.00000 11.71203
-125.00000 7.43470
-125.00000 11.93368
-124.00000 8.73525
-124.00000 7.23886
-123.00000 8.91606
-123.00000 7.19680
-123.00000 10.69766
-123.00000 7.75807
-123.00000 10.62275
-122.00000 8.42779
-122.00000 8.01548
-122.00000 9.18641
-122.00000 8.00275
-122.00000 7.25097
-121.00000 8.00625
-121.00000 7.73302
-121.00000 8.33637
-121.00000 7.05740
-121.00000 9.37223
-120.00000 6.19740
-120.00000 7.67970
-120.00000 8.88046
-119.00000 6.15293
};
\addlegendentry{CE Level 1}

\addplot [only marks, mark=*, mark size=1.5pt, draw=color5, fill=color5, colormap/viridis]
table{%
x                      y
-137.00000 14.79844
-137.00000 19.53369
-136.00000 9.72451
-135.00000 19.16831
-133.00000 11.14180
-133.00000 13.17793
-133.00000 13.26506
-132.00000 13.00898
-132.00000 11.04822
-132.00000 10.72677
-132.00000 18.67828
-132.00000 17.63324
-131.00000 19.30826
-131.00000 18.27587
-131.00000 11.42717
-131.00000 14.83598
-131.00000 16.23786
-131.00000 16.73903
-131.00000 10.84045
-131.00000 11.22467
-131.00000 18.61850
-131.00000 7.89095
-131.00000 12.34875
-130.00000 14.52603
-130.00000 16.25924
-130.00000 19.98432
-130.00000 12.74129
-130.00000 11.48145
-129.00000 20.16765
-129.00000 17.78946
-129.00000 10.64937
-129.00000 11.99991
-128.00000 12.01975
-128.00000 11.91607
-128.00000 11.20879
-128.00000 10.43471
-126.00000 10.55384
-126.00000 16.87997
-126.00000 12.31109
-126.00000 14.28163
-125.00000 12.94446
};
\addlegendentry{CE Level 2}
\end{axis}

\end{tikzpicture}
    \caption{\small Energy consumption of an \gls{nbiot} modem by sending a 5-byte payload (18 byte overhead) relative to the observed received power (\gls{rsrp}). Adapted  from~\cite{leenders2020experimental}. 
    }
    \label{fig:totalenergynbiot}
\end{figure}

\subsection{Energy Efficiency}
\label{sec:evaluation_energy}

In order to compare the energy efficiency of \gls{lorawan} and \gls{nbiot}, the energy required to transmit one payload byte \(E_B\) has been determined. Energy measurements only consider the energy used by the \gls{iot} transceiver. 

\lorawan

The measured power profile of a \gls{lorawan} message is depicted in Fig.~\ref{fig:currentdrawlora}. \new{In this example a payload of \SI{5}{\byte} is sent with a header of \SI{13}{\byte} (\gls{sf} 9).} 
The model of~\cite{callebaut2021art} combines the measured power consumption in each state and the theoretical air time of a message to compute the total energy consumption of a 
message. Several states can be observed: transmission of the packet, two processing slots, and two receive windows for receiving downlink messages. Fig.~\ref{fig:energyperbyte} shows the energy per payload byte for different payloads sizes. The payload needs to be divided over multiple messages when the payload size exceeds the maximum message length.

\nbiot
A typical experimental power consumption measurement is shown in Fig.~\ref{fig:currentdrawnbiot}. Several modem states can be observed: network search and join, package send, connected mode \gls{drx}, \gls{edrx} and power saving mode. 

During the network search and join mechanism, the modem boots and tries to connect to the network. When connecting to the network, the modem negotiates several network parameters (e.g., \gls{ce} level, timer values, etc.). After the node has successfully joined the network, the package is sent. 
In this example, five bytes of data encapsulated in a \gls{udp} packet, resulting in a total \gls{nbiot}-payload of \SI{23}{\byte}, is sent.  
After the message is sent, the modem enters \gls{cdrx}. The radio circuitry stays active for a predetermined period, so downlink messages can be received. 
If \gls{edrx} is supported by both the network provider and the \gls{nbiot} node, the modem is able to support periodical downlink communication. To conserve energy, the modem is put to sleep between downlink windows: \gls{ptw}. After a set of \gls{edrx} cycles, the modem enters its lowest possible energy state: \gls{psm}. Herein, the modem is put to sleep and is not available for any network communication. When the specified \gls{psm} timer runs out, the modem reconnects to the network. It is also possible to prematurely reinstate the network communication based on interrupts. 

Higher \gls{ce} levels amount to a larger energy consumption of a \gls{nbiot} node. After measuring the \gls{rsrp}, the node negotiates the appropriate \gls{ce} level with the network. This decision directly influences energy consumption for the longevity of the connection. 
This procedure is experimentally validated in Fig.~\ref{fig:totalenergynbiot}. The total energy consumption is depicted, with respect to the \gls{rsrp}: the \gls{rf} signal quality indicator on the \gls{nbiot} node. One can clearly see that a lower RSRP will result in a higher \gls{ce} level being selected. Moreover, the results clearly indicate that a node transmitting in \gls{ce} level~2 can use up to four times more energy than a node transmitting in \gls{ce} level~0. 

\gls{nbiot} supports a maximum of \SI{1600}{\byte} to be sent in a single packet. The effective payload size also influences the energy efficiency of \gls{nbiot}. As fixed energy costs, such as \gls{cdrx} and \gls{edrx}, are equal for smaller and larger payload sizes, the energy per byte  (\(E_B\)) will be lower for longer packets (Fig.~\ref{fig:energyperbyte}). Notably, \gls{nbiot} \new{(at \gls{ce} level 2)} is more energy efficient than \gls{lorawan} (at \gls{sf} 12) when the payload size exceeds \new{\SI{240}{\byte}}. For payloads larger than \new{\SI{240}{\byte}}, \gls{nbiot} becomes progressively \liesbetok{kunnen we hier nog iets simpel kwantitatief invoegen, bv progressively and up to 8 times?} and up to 7 times more energy efficient than \gls{lorawan} with larger payloads. 
The energy impact of rising \gls{ce} levels, however, is reduced when transmitting large payloads. 

\subsection{Latency}
Latency in \gls{lorawan} and \gls{nbiot} was evaluated, hereby focusing on uplink latency. 

\lorawan
The uplink latency is limited by the allowed air time or, equivalently, duty cycle. The uplink latency for payloads sizes of 1600 bytes ranges from \SI{42}{\minute} to \SI{68}{\hour} (only the default and mandatory bands are used, i.e., 1\% duty cycle). As the maximum message size is constrained by the spreading factor, the payload needs to be split up in multiple messages in order to send the full payload, increasing the uplink latency. The total duty cycle is limited by the supported operating bands of the network.

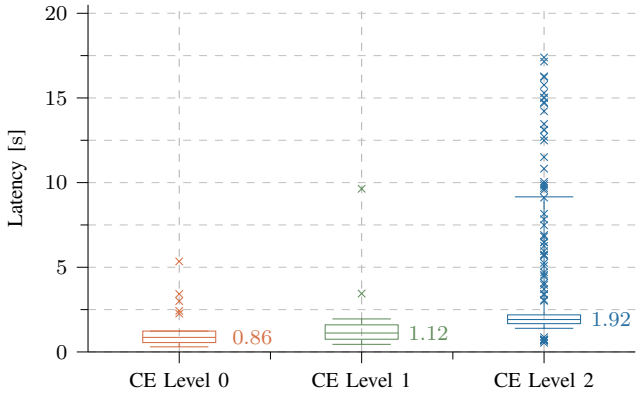
\begin{figure}[tbp]
    \centering
\begin{tikzpicture}

\begin{axis}[
axis lines* = {left},
tick align=outside,
tick pos=left,
width = \linewidth,
height = 0.7\linewidth,
x grid style={white!69.01960784313725!black},
grid,
grid style={help lines,color=gray!50, dashed},
xmajorgrids,
xmin=0.50000, xmax=3.50000,
xtick style={color=black},
xtick={1.00000,2.00000,3.00000},
xticklabels={CE Level 0,CE Level 1, CE Level 2},
minor tick num=1,
yminorgrids,
ymajorgrids,
ylabel={Latency [\si{\second}]},
ymin=0, ymax=20.5,
ytick style={color=black},
boxplot/draw direction=y,
label style={font=\footnotesize},
tick label style={font=\footnotesize}
]

\addplot[mark = x, mark options = {mark color=grey},
boxplot prepared={
    lower whisker=0.301, 
    lower quartile=0.55450,
    median=0.85900,
    upper quartile=1.23100, 
    upper whisker=1.231,
    box extend=0.4}, 
    color=colorDarkCE0,
]
coordinates{
(1, 2.26300)
(1, 3.00300)
(1, 5.34500)
(1, 2.42600)
(1, 3.42200)
}
node[right,font=\footnotesize] at
(boxplot box cs: \boxplotvalue{median},1.1)
{\pgfmathprintnumber{\boxplotvalue{median}}};;

\addplot[mark = x, mark options = {mark color=grey},
boxplot prepared={
    lower whisker=0.448, 
    lower quartile=0.7505,
    median=1.117,
    upper quartile=1.6, 
    upper whisker=1.95,
    box extend=0.4}, 
    color=colorDarkCE1
]
coordinates{
(2, 9.63600)
(2, 3.44700)
}
node[right,inner sep=1pt,font=\footnotesize] at
(boxplot box cs: \boxplotvalue{median},1.1)
{\pgfmathprintnumber{\boxplotvalue{median}}};;

\addplot[mark = x, mark options = {mark color=grey},
boxplot prepared={
    lower whisker=1.395, 
    lower quartile=1.6715,
    median=1.915,
    upper quartile=2.1895, 
    upper whisker=9.16,
    box extend=0.4}, 
    color=colorDarkCE2
]
coordinates{
(3.00000 ,0.72500)
(3.00000 ,0.70400)
(3.00000 ,0.67100)
(3.00000 ,0.70900)
(3.00000 ,0.69400)
(3.00000 ,0.73600)
(3.00000 ,0.65700)
(3.00000 ,0.72200)
(3.00000 ,0.54200)
(3.00000 ,0.87100)
(3.00000 ,5.95000)
(3.00000 ,3.01900)
(3.00000 ,4.69900)
(3.00000 ,13.11500)
(3.00000 ,12.70200)
(3.00000, 8.15800)
(3.00000, 13.10800)
(3.00000, 5.68100)
(3.00000, 15.76600)
(3.00000, 6.81500)
(3.00000, 16.27500)
(3.00000, 10.82400)
(3.00000, 3.96200)
(3.00000, 3.70600)
(3.00000, 15.28800)
(3.00000, 3.43400)
(3.00000, 9.91600)
(3.00000, 3.38100)
(3.00000, 9.11200)
(3.00000, 9.67800)
(3.00000, 16.22900)
(3.00000, 5.25500)
(3.00000, 12.48700)
(3.00000, 21.67800)
(3.00000, 7.83200)
(3.00000, 14.68500)
(3.00000, 10.04800)
(3.00000, 13.46300)
(3.00000, 14.74700)
(3.00000, 7.48200)
(3.00000, 5.72400)
(3.00000, 6.26300)
(3.00000, 4.07000)
(3.00000, 6.90300)
(3.00000, 6.48400)
(3.00000, 14.21600)
(3.00000, 4.46700)
(3.00000, 6.46900)
(3.00000, 3.07000)
(3.00000, 4.55700)
(3.00000, 5.75500)
(3.00000, 15.02500)
(3.00000, 23.52000)
(3.00000, 5.08600)
(3.00000, 14.99400)
(3.00000, 17.15000)
(3.00000, 17.39200)
(3.00000, 9.81500)
(3.00000, 11.50900)
(3.00000 ,9.56200)
}
node[right,inner sep=1pt,font=\footnotesize] at
(boxplot box cs: \boxplotvalue{median},1.1)
{\pgfmathprintnumber{\boxplotvalue{median}}};;
\end{axis}

\end{tikzpicture}
    \caption{\small Boxplot of measured \gls{nbiot} latency when sending a \new{packet}: \SI{14}{\byte} payload, \gls{ce} level 2.%
    } 
    \label{fig:nbiotlatency}
    
\end{figure}

\nbiot
According to \gls{nbiot} specifications, the latency should be kept under \SI{10}{\second}. This was experimentally validated by sending 1236 packages 
in varying signal conditions. 
The results, depicted in Fig.~\ref{fig:nbiotlatency}, show that both \gls{ce} levels~0 and~1 keep latency within the \SI{10}{\second} maximum. Packets transmitted in \gls{ce} level 1 only have a slightly larger latency than packets transmitted in \gls{ce} level~0.  When transmitting on \gls{ce} level~2, however, poor signal conditions can cause latency to reach up to \SI{20}{\second}. The median latency increases and more extreme latency outliners occur. In \gls{nbiot}, latency is barely influenced by payload size. 

\section{Conclusions}

Considering the predominant requirements for an \gls{lpwan} node, our assessment has highlighted the complementarities and potential synergies of \gls{lorawan} and \gls{nbiot} technologies. \new{While for many basic \gls{iot} applications this leads to one preferred technology, implementing a \gls{multi-rat} scheme in more dynamic use cases can greatly improve a node's functionality and energy efficiency.} Based on our study and experimental validation, we demonstrate the potential of a \gls{multi-rat} (\gls{lorawan}/\gls{nbiot}) solution from the perspective of different \gls{iot} requirements, while focusing on an optimal energy trade-off. 

\subsection{Potential \gls{multi-rat} Gains}

\paragraph{Energy Efficient Operation for Variable Payload Sizes} \gls{iot} use cases with varying payload sizes benefit from a \gls{multi-rat} approach. By implementing \gls{nbiot}, messages up to \SI{1600}{\byte} can be sent, while still enabling extremely low-power messages with small payloads over \gls{lorawan}. Considering the case of smart city surveillance, whereby a node would send hourly sensor data (e.g., sound level), and more extensive data when certain thresholds are met (e.g. loud noises). 
The \gls{multi-rat} solution would select \gls{lorawan} for hourly sensor updates (\new{based on existing use cases~\cite{callebaut2021art}:} \SI{16}{\byte}, 97\% of packages), and \gls{nbiot} for sending more elaborate sensor recordings for classification (assumed \SI{1600}{\byte}, occurring for example 5 times per week or 3\% of sent messages). This would result in \SI{23}{\joule} transmit energy per week, a reduction of times 15 with respect to only using  \gls{nbiot} and of times 4 when only using \gls{lorawan}. \new{To make this more tangible, this would mean that the device could, in ideal conditions, operate for 30 years on a single smartphone battery (2500 mAh).} 
\Gls{lorawan} would require the large payloads being split into 32 messages, which would clearly also introduce a latency penalty.  

\paragraph{Guaranteeing Timely Delivery for Latency-Critical Messages} Depending on network coverage and network load, latency can be optimized by spreading communications over multiple \gls{iot} technologies. When low energy consumption is critical and the \gls{nbiot} chipset is in \gls{psm}, a faster wake-up can be achieved with \gls{lorawan}. On the other hand, larger payloads can be sent more rapidly with \gls{nbiot}. This could be important for monitoring medical grade parameters (e.g., heart rhythm or fall detection).

\paragraph{\new{Redundant Networking and Improving Service Area}} \new{By combining multiple \gls{iot} technologies, the effective service area of a \gls{multi-rat} solution will be extended to the area of all \gls{iot} technologies on-board. When coverage is not provided by one \gls{iot} technology, another can step in.   The service area can be privately extended for \gls{lorawan} by deploying private gateways.}

\paragraph{Improving \gls{qos}} By enabling an \gls{iot} device to operate on both \gls{lorawan} and \gls{nbiot}, the optimal \gls{qos} for any message can be chosen: increasing robustness and reliability. Periodic `alive' messages do not need high \gls{qos}, yet more important messages containing sensitive data (e.g., temperature tracking on track and trace applications) do need high \gls{qos}. \Gls{lorawan} should be used for the periodic, low \gls{qos}, messages, saving energy. Important messages can be sent through the \gls{nbiot} network: featuring high \gls{qos}.

\subsection{Potential \gls{multi-rat} Drawbacks}
\paragraph{Device Footprint} By including multiple \gls{iot} technologies on one device, multiple modems will need to be on-board. This increases the space needed for wireless interfaces by both the applicable modem size and the appropriate antenna size. The board space occupied by multiple modems, can be improved by efficiently using both sides of a \gls{pcb}. Multiple antennas can be combined in multi-band antennas.

\paragraph{Device Cost} As multiple modems are included, hardware and network subscription costs will rise. However, by optimizing energy consumption, costs can be saved by not requiring manual intervention for battery replacements.
\paragraph{Computational Overhead} To dynamically switch between \gls{iot} technologies, some computational overhead will be required. The energy savings from implementing a \gls{multi-rat} platform, however, the energy savings of \gls{multi-rat} outweighs the energy consumption of the required additional computations.~\cite{karl2007protocols}. 

\color{black}
\subsection{Conclusions}
In many cases, the prioritization of strict \gls{iot} requirements is not an easy feat. For example, when an application requires both small and larger data transfers, the choice of \gls{iot} technology will affect the energy consumption, latency, etc., of both types of messages. 
The \gls{multi-rat} solution presented in this paper allows to dynamically adapt the stated \gls{iot} properties priorities (and thus \gls{iot} technology), depending on the most prominent \gls{iot} requirement and current context. This in particular benefits energy consumption that can be reduced by impressive factors. \new{The gathered empirical data can help to select either the most appropriate connectivity solution and contribute to a behavioural model of the latency and power consumption, used to develop \gls{multi-rat} dynamic operation procedures: optimizing efficiency and effectivity.}

{
\footnotesize

\bibliographystyle{IEEEtranNMod}
\bibliography{sources.bib}{}
}

\newpage
\vfill%
{\footnotesize%
\begin{IEEEbiographynophoto}{Guus Leenders}%
Guus Leenders received his master’s degree summa cum laude in engineering technology at KU Leuven in 2015. He is member of the Dramco research group. There, he is involved with numerous projects in IoT. His interests are Internet of Things and embedded systems.
\end{IEEEbiographynophoto}
\begin{IEEEbiographynophoto}{Gilles Callebaut}
Gilles graduated summa cum laude in 2016 and received the M.Sc. degree in engineering technology at KU Leuven. He is currently a member of the Dramco reserach group. His interests are Machine Type Communication, Internet of Things, embedded systems and everything mobile.
\end{IEEEbiographynophoto}
\begin{IEEEbiographynophoto}{Geoffrey Ottoy}
Geoffrey is a researcher on Internet of Things applications. In 2013 he received his Ph.D. in Electrical Engineering from KU Leuven. His interests are indoor localization and low-power embedded systems.
\end{IEEEbiographynophoto}
\begin{IEEEbiographynophoto}{Liesbet Van der Perre}
Liesbet Van der Perre received her PhD degree in Electrical Engineering from KU Leuven, Belgium. She is  professor at Dramco, KU Leuven. Her research interest are energy efficient wireless communication and embedded systems.
\end{IEEEbiographynophoto}%
\begin{IEEEbiographynophoto}{Lieven De Strycker}
Lieven De Strycker is professor at the Faculty of Engineering Technology, Department of Electrical Engineering, KU Leuven. He joined the Engineering Technology department of the Catholic University College Ghent, where he founded the Dramco research group at KU Leuven. 
\end{IEEEbiographynophoto}\vfill}
\null
\clearpage

\setcounter{figure}{0} 
\pagenumbering{gobble}

\begin{figure}[p!]
    \centering
    \begin{tikzpicture}
\tkzKiviatDiagram[scale=0.3,label distance=.2cm,
        radial  = 6,
        gap     = 1,  
        label space=2.2,
        label style/.append style={font=\footnotesize},
        lattice = 5.5]{{Energy\\Efficiency}, {Coverage}, {Payload\\Size},{Latency\\Performance}, {Quality of\\Service}, {Cost \\Efficiency}}
\tkzKiviatLine[thick,color=colorLoRa,mark=none,   fill=colorLoRa!20,opacity=.5](4.5,3,2,2,3,4.5) 
\tkzKiviatLine[thick,color=colorNBIoT,                      fill=colorNBIoT!20,opacity=.5](3.5,4,4.5,4,4.5,3) 
\tkzKiviatLine[dashed, thick,color=colorMultiRAT,                      fill=colorMultiRAT!0,opacity=.5](4.5,5,5,4,4.5,4) 
 
\end{tikzpicture}

\tkzKiviatLegend{LoRaWAN}{NB-IoT}{Multi-RAT LoRaWAN, NB-IoT}
    \caption{\small Comparative study of the explored \gls{iot} network technologies, presenting the main \gls{iot} feature requirements.  }
\end{figure}
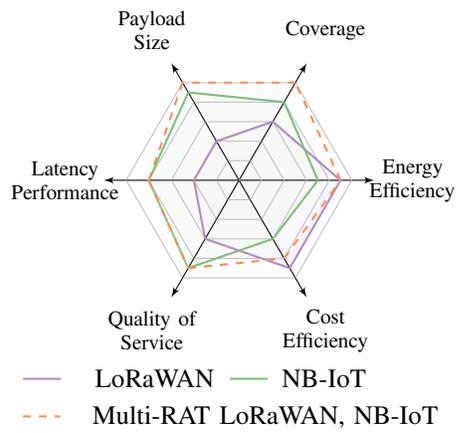
\clearpage
\begin{figure*}[p!]
	\centering
	\begin{subfigure}[t]{0.49\linewidth}
    \centering
    \input{img/lora_energyplot.tex}
    \caption{\small Measured power consumption of a \gls{lora} node (at \gls{sf} 9): (1) transmit, (2) processing, (3) first receive window, (4) processing, (5) second receive window. Network registration is omitted, as this is not obligatory in \gls{lorawan}.}  
	\end{subfigure}\hfill
	\begin{subfigure}[t]{0.49\linewidth}
    \centering
    \input{img/nbiot_energyplot}
    \caption{\small Measured power consumption of a \gls{nbiot} node (at \gls{ce} level~0): (1) network search and join, (2) package sending, (3) \acrlong{cdrx}, (4) \acrlong{edrx}, (5) \acrlong{psm}. Network registration is obligatory and is only performed once (as long as power is maintained). }
	\end{subfigure}
	\caption{\small Experimentally determined power consumption of \gls{lora} and \gls{nbiot} side by side. Note the large difference in timing, resulting in a larger energy consumption for \gls{nbiot}. }
\end{figure*}
\clearpage
\begin{figure}[p!]
    \centering
    \input{img/lora_energy_per_payload_byte.tex}
    \caption{\small Energy consumption per byte comparison between \gls{lorawan} and \gls{nbiot}. The difference in energy consumption is most noticeable when sending small payloads. Over \new{\SI{240}{\byte}}, \gls{nbiot} \new{(\gls{ce} level 2)} is more energy efficient per byte, with respect to \gls{lorawan} (\gls{sf} 12). 
    }
\end{figure}
\clearpage
\begin{figure}[p!]
    \centering
\begin{tikzpicture}

\colorlet{color1}{colorCE0}
\colorlet{color4}{colorCE1}
\colorlet{color5}{colorCE2}

\begin{axis}[%
axis lines* = {left},
tick align=outside,
tick pos=left,
width = \linewidth,
height = 0.6\linewidth,
separate axis lines,
x grid style={white!69.01960784313725!black},
grid,
grid style={help lines,color=gray!50, dashed},
xmin=-140.77178, xmax=-66.62358,
xtick style={color=black},
y grid style={white!69.01960784313725!black},
ytick={0,10,20},
xtick={-140,-120,-100,-80,-60},
ymin=2.89056, ymax=22.70769,
ytick style={color=black},
legend style={draw=none, nodes={scale=0.7, transform shape}},
xlabel={RSRP [\si{\dBm}]},
ylabel={Energy [\si{\joule}]},
label style={font=\footnotesize},
tick label style={font=\footnotesize},
minor tick num=1,
yminorgrids,
ymajorgrids,
xminorgrids
]
\addplot [only marks, mark=*, mark size=1.5pt, draw=color1, fill=color1, colormap/viridis]
table{%
x                      y
-119.00000 6.62973
-118.00000 7.45417
-118.00000 7.53283
-117.00000 8.23190
-116.00000 6.10285
-116.00000 7.13637
-116.00000 5.73844
-115.00000 7.04443
-115.00000 6.62073
-114.00000 6.36626
-114.00000 6.86505
-111.00000 5.72806
-111.00000 6.14779
-110.00000 5.86173
-110.00000 6.22797
-109.00000 5.33608
-108.00000 6.47111
-107.00000 5.70141
-107.00000 5.73046
-105.00000 4.75373
-105.00000 5.41399
-104.00000 6.23227
-100.00000 5.71132
-99.00000 5.16229
-98.00000 6.12374
-95.00000 5.29644
-94.00000 5.75344
-89.00000 5.19242
-80.00000 6.62887
-78.00000 5.17348
-77.00000 4.20855
-77.00000 4.95602
-75.00000 4.65286
-75.00000 5.29805
-75.00000 5.63507
-74.00000 4.79664
-74.00000 5.20403
-73.00000 5.64619
-73.00000 4.56572
-73.00000 5.49128
-72.00000 4.48788
-70.00000 5.95012
};
\addlegendentry{CE Level 0}

\addplot [only marks, mark=*, mark size=1.5pt, draw=color4, fill=color4, colormap/viridis]
table{%
x                      y
-130.00000 14.10063
-129.00000 10.85486
-129.00000 7.44360
-128.00000 14.38983
-128.00000 11.32614
-127.00000 10.30209
-127.00000 9.63297
-127.00000 8.33477
-126.00000 9.66211
-126.00000 8.23243
-126.00000 8.10336
-126.00000 8.91433
-126.00000 8.26592
-126.00000 12.85321
-126.00000 10.29418
-125.00000 7.15848
-125.00000 10.58148
-125.00000 11.87369
-125.00000 11.71203
-125.00000 7.43470
-125.00000 11.93368
-124.00000 8.73525
-124.00000 7.23886
-123.00000 8.91606
-123.00000 7.19680
-123.00000 10.69766
-123.00000 7.75807
-123.00000 10.62275
-122.00000 8.42779
-122.00000 8.01548
-122.00000 9.18641
-122.00000 8.00275
-122.00000 7.25097
-121.00000 8.00625
-121.00000 7.73302
-121.00000 8.33637
-121.00000 7.05740
-121.00000 9.37223
-120.00000 6.19740
-120.00000 7.67970
-120.00000 8.88046
-119.00000 6.15293
};
\addlegendentry{CE Level 1}

\addplot [only marks, mark=*, mark size=1.5pt, draw=color5, fill=color5, colormap/viridis]
table{%
x                      y
-137.00000 14.79844
-137.00000 19.53369
-136.00000 9.72451
-135.00000 19.16831
-133.00000 11.14180
-133.00000 13.17793
-133.00000 13.26506
-132.00000 13.00898
-132.00000 11.04822
-132.00000 10.72677
-132.00000 18.67828
-132.00000 17.63324
-131.00000 19.30826
-131.00000 18.27587
-131.00000 11.42717
-131.00000 14.83598
-131.00000 16.23786
-131.00000 16.73903
-131.00000 10.84045
-131.00000 11.22467
-131.00000 18.61850
-131.00000 7.89095
-131.00000 12.34875
-130.00000 14.52603
-130.00000 16.25924
-130.00000 19.98432
-130.00000 12.74129
-130.00000 11.48145
-129.00000 20.16765
-129.00000 17.78946
-129.00000 10.64937
-129.00000 11.99991
-128.00000 12.01975
-128.00000 11.91607
-128.00000 11.20879
-128.00000 10.43471
-126.00000 10.55384
-126.00000 16.87997
-126.00000 12.31109
-126.00000 14.28163
-125.00000 12.94446
};
\addlegendentry{CE Level 2}
\end{axis}

\end{tikzpicture}
    \caption{\small Energy consumption of an \gls{nbiot} modem by sending a 5-byte payload (18 byte overhead) relative to the observed received power (\gls{rsrp}). Adapted  from~\cite{leenders2020experimental}. 
    }
\end{figure}
\clearpage
\begin{figure}[p!]
    \centering
\begin{tikzpicture}

\begin{axis}[
axis lines* = {left},
tick align=outside,
tick pos=left,
width = \linewidth,
height = 0.7\linewidth,
x grid style={white!69.01960784313725!black},
grid,
grid style={help lines,color=gray!50, dashed},
xmajorgrids,
xmin=0.50000, xmax=3.50000,
xtick style={color=black},
xtick={1.00000,2.00000,3.00000},
xticklabels={CE Level 0,CE Level 1, CE Level 2},
minor tick num=1,
yminorgrids,
ymajorgrids,
ylabel={Latency [\si{\second}]},
ymin=0, ymax=20.5,
ytick style={color=black},
boxplot/draw direction=y,
label style={font=\footnotesize},
tick label style={font=\footnotesize}
]

\addplot[mark = x, mark options = {mark color=grey},
boxplot prepared={
    lower whisker=0.301, 
    lower quartile=0.55450,
    median=0.85900,
    upper quartile=1.23100, 
    upper whisker=1.231,
    box extend=0.4}, 
    color=colorDarkCE0,
]
coordinates{
(1, 2.26300)
(1, 3.00300)
(1, 5.34500)
(1, 2.42600)
(1, 3.42200)
}
node[right,font=\footnotesize] at
(boxplot box cs: \boxplotvalue{median},1.1)
{\pgfmathprintnumber{\boxplotvalue{median}}};;

\addplot[mark = x, mark options = {mark color=grey},
boxplot prepared={
    lower whisker=0.448, 
    lower quartile=0.7505,
    median=1.117,
    upper quartile=1.6, 
    upper whisker=1.95,
    box extend=0.4}, 
    color=colorDarkCE1
]
coordinates{
(2, 9.63600)
(2, 3.44700)
}
node[right,inner sep=1pt,font=\footnotesize] at
(boxplot box cs: \boxplotvalue{median},1.1)
{\pgfmathprintnumber{\boxplotvalue{median}}};;

\addplot[mark = x, mark options = {mark color=grey},
boxplot prepared={
    lower whisker=1.395, 
    lower quartile=1.6715,
    median=1.915,
    upper quartile=2.1895, 
    upper whisker=9.16,
    box extend=0.4}, 
    color=colorDarkCE2
]
coordinates{
(3.00000 ,0.72500)
(3.00000 ,0.70400)
(3.00000 ,0.67100)
(3.00000 ,0.70900)
(3.00000 ,0.69400)
(3.00000 ,0.73600)
(3.00000 ,0.65700)
(3.00000 ,0.72200)
(3.00000 ,0.54200)
(3.00000 ,0.87100)
(3.00000 ,5.95000)
(3.00000 ,3.01900)
(3.00000 ,4.69900)
(3.00000 ,13.11500)
(3.00000 ,12.70200)
(3.00000, 8.15800)
(3.00000, 13.10800)
(3.00000, 5.68100)
(3.00000, 15.76600)
(3.00000, 6.81500)
(3.00000, 16.27500)
(3.00000, 10.82400)
(3.00000, 3.96200)
(3.00000, 3.70600)
(3.00000, 15.28800)
(3.00000, 3.43400)
(3.00000, 9.91600)
(3.00000, 3.38100)
(3.00000, 9.11200)
(3.00000, 9.67800)
(3.00000, 16.22900)
(3.00000, 5.25500)
(3.00000, 12.48700)
(3.00000, 21.67800)
(3.00000, 7.83200)
(3.00000, 14.68500)
(3.00000, 10.04800)
(3.00000, 13.46300)
(3.00000, 14.74700)
(3.00000, 7.48200)
(3.00000, 5.72400)
(3.00000, 6.26300)
(3.00000, 4.07000)
(3.00000, 6.90300)
(3.00000, 6.48400)
(3.00000, 14.21600)
(3.00000, 4.46700)
(3.00000, 6.46900)
(3.00000, 3.07000)
(3.00000, 4.55700)
(3.00000, 5.75500)
(3.00000, 15.02500)
(3.00000, 23.52000)
(3.00000, 5.08600)
(3.00000, 14.99400)
(3.00000, 17.15000)
(3.00000, 17.39200)
(3.00000, 9.81500)
(3.00000, 11.50900)
(3.00000 ,9.56200)
}
node[right,inner sep=1pt,font=\footnotesize] at
(boxplot box cs: \boxplotvalue{median},1.1)
{\pgfmathprintnumber{\boxplotvalue{median}}};;
\end{axis}

\end{tikzpicture}
    \caption{\small Boxplot of measured \gls{nbiot} latency when sending a \new{packet}: \SI{14}{\byte} payload, \gls{ce} level 2.%
    } 
\end{figure}
\clearpage
\end{document}